\documentclass[prl,twocolumn,amsfonts,bibnotes]{revtex4}

\usepackage{amsmath}
\usepackage{bbm}
\usepackage{hyperref}
\usepackage{graphicx}
\usepackage{pifont}

\newcommand{\mc}{\mathcal}
\newcommand{\ket}[1]{|\, #1\rangle}
\newcommand{\bra}[1]{\langle #1\,|}

\def\tab{\>\>\>\>}
\def\squarebox#1{\hbox to #1{\hfill\vbox to #1{\vfill}}}
\def\qed{\hspace*{\fill}\vbox{\hrule\hbox{\vrule\squarebox{.667em}\vrule}\hrule}}
\newenvironment{proof}{\begin{trivlist}\item[]{\bf Proof:}}{\qed \end{trivlist}}
\newtheorem{theo}{Theorem}
\newtheorem{lem}{Lemma}

\begin{document}

\title{Quantum information with single fermions:\\
teleportation and fermion-boson entanglement conversion}

\author{ O. Morgenshtern }
\affiliation{ Department of Physics and Astronomy, Tel Aviv University, Tel
Aviv 69978, Israel.
       }
\author{ B. Reznik }
\affiliation{ Department of Physics and Astronomy, Tel Aviv University, Tel
Aviv 69978, Israel. }
\author{ I. Zalzberg }
\affiliation{ Department of Physics and Astronomy, Tel Aviv University, Tel
Aviv 69978, Israel.
       }

\date{\today}

\begin{abstract}

While single boson entanglement is  known to be equivalent to standard non-identical particle entanglement, the use of fermionic entanglement 
is constrained by a parity superselection rule.
Nevertheless we show that within the framework of second quantization, 
where local modes play the role of qubits, the analogy with ordinary entanglement holds: two entangled spinless fermion modes can be used to teleport a single local fermion mode and to perform standard dense coding. Entanglement concentration and dilution are possible for fermions. We clarify the meaning of our results by  discussing the connection with quantum reference frames and  conversion between spinless fermion entanglement and spinless boson entanglement.
\end{abstract}
\maketitle




The principle of causality, together with the anti-commutativity of fermionic modes creation/annihilation operators impose
a restriction on any fermionic interaction;
fermions can only be created or destroyed in pairs and  are hence known
to obey a parity superselection rule (SSR)  \cite{www,Aharonov-Susskind}.
Therefore, while single (spinless) boson entanglement is known to be equivalent to standard non-identical particle entanglement\cite{Tan,Hardy}, the situation is different for single (spinless) fermion entanglement.
Since set of allowed  local operations is constrained by parity SSR
the nature of fermionic entanglement  has a different stand and may seem perhaps less interesting from the point of view of quantum information.


In the present work we reexamine the problem within the framework of 
second quantization.  The basic objects are then local mode number states
which are formally analogous to qubits. Entanglement is considered between sets of local modes \cite{Zanardi,Mari-Carmen,particle-entanglement}, and
the simplest possible form
of spinless fermionic bipartite entanglement, is equivalent to having a  single fermion in a state of a superposition at two remote locations, that  in this framework is described by a two local mode state  $|1,0\rangle+ |0,1\rangle$.
We shall refer to the latter as an `e-mode'.

Surprisingly, we  find that spinless fermion entanglement can in fact be used in ``canonic" quantum protocols, and that teleportation, dense coding and distillations,  are possible in such a ``purely fermionic" system.
The standard protocols can be formulated in a unified form applicable for either fermions or bosons:
\newline
1. A single mode can be teleported using
a single shared e-mode and sending two classical bits.  (Fig. 1).
\newline
2. Two classical bits can be transmitted using one shared e-mode + the ability to send one quantum mode from Alice to Bob (dense coding).
\newline 3.
 Concentration and dilution procedures of e-modes are possible with fermions (bosons).

Since spinless bosons are not constrained by SSR, a two dimensional subspace of a single bosonic modes can be used to represent a qubit, and bosonic e-modes are equivalent to non-identical particle entanglement .
Nevertheless, the above statements are not obvious for fermionic entanglement, since, as we prove, the conversion between a single fermion e-mode to a single bosonic e-mode and vise versa is not possible.

Further insights into the problem can be gained by
considering the role of quantum reference frames 
which have been recently been studied in connection 
to SSR and implications in quantum information\cite{Bartlett-rmp}. It has been realized that much of the restrictions due to SSR can be lifted by introducing a suitable quantum reference frame\cite{Aharonov-Susskind}.
On the other hand, in the absence of a suitable reference system, SSR
restricts the set of local operations and gives rise to a new resource, Superselection induced
Variance (SiV), which qualitatively amounts to local uncertainty in a SSR conserved quantity~\cite{Schuch}. It has been shown that SiV can be used for tasks which otherwise are not possible\cite{Vesrstraete}.

\begin{figure}[h]
\centering
\includegraphics[width=70mm]{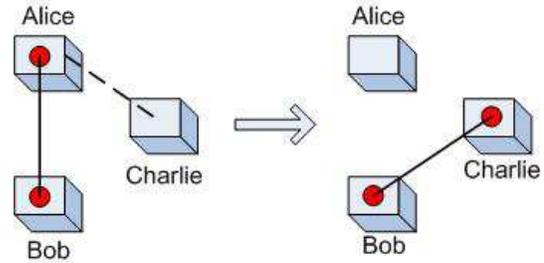}
\caption{Single fermion mode teleportation:  Alice's mode, which is part of a general fermion state between Alice and Bob (solid line),
is teleported to Charlie using one fermionic e-mode (dashed line)
between Alice and Charlie.  This process cannot be carried out using only bosonic entanglement. }
\end{figure}

We show that fermion mode teleportation can in fact be carried out using pure spinless bosonic entanglement, provided that we supplement it with a suitable reference frame. Such a reference frame has to carry a nonzero SiV resource however it may consist of a non-entangled bi-partite mixed fermionic state.
In fact in the presence of such a reference frame, bosonic and fermionic entanglement become convertible with unit probability while the reference frame acts as an catalyzer; it is not consumed in the conversion process. From this point of view, the curious feature of a single fermion e-mode entanglement, is that a single fermion e-mode seems to carry simultaneously entanglement and its own Siv, which together are enough to enable a deterministic fermion mode teleportation.

Let us begin by considering Bell states of two fermion modes. $\phi^\pm=\ket{00} \pm \ket{11}$ and $\psi^\pm=\ket{01} \pm \ket{10}$ have a parity number of '1' and '-1' respectively. The parity operator is $\hat P=\displaystyle\prod_{i}(-1)^{n_i}$, where $n_i$ is the number of fermions in the i'th mode. It is easy to see that the above
states are eigenstates of the Hermitian ''Bell operators'':
$\mc O_{1} = a^{\dagger}b^{\dagger} + ba$ and $\mc O_{2} = a^{\dagger}b + b^{\dagger}a$, $[O_i,P]=0$,
where $a$ and $b$ are the relevant mode operators.
SSR prohibits a superposition of Bell states with
different parity. Nevertheless it is possible by means of a local operation
(on say mode $a$) to map any Bell state to another.
To see this, we note that phase inversion is trivially possible. Local bit flip would result in a violation of the parity SSR. In order to overcome this restriction, we add locally an ancillary fermion mode in the $\ket{0}$ state, and perform the following local two-mode transformations:
\begin{eqnarray}\label{eq:BitFlip}
\ket{00}\Rightarrow\ket{11}, \tab
\nonumber\ket{11}\Rightarrow\ket{00}, \tab
\nonumber\ket{01}\Rightarrow\ket{10}, \tab
\nonumber\ket{10}\Rightarrow\ket{01}
\end{eqnarray}
The above can be used to transform:
\begin{eqnarray}\label{eq:dense-coding}
|0\rangle|\psi^\pm\rangle \Rightarrow  |1\rangle|\phi^\pm\rangle \tab
|0\rangle|\phi^\pm\rangle \Rightarrow  |1\rangle|\psi^\pm\rangle
\end{eqnarray}
It is now clear that Alice can use a single fermion e-mode for the purpose of dense-coding.
 She adds a local ancillary fermion mode, (which is assumed to be a free resource),
  and using the above map she can encode two bits in the usual manner by locally rotating the shared state, say $\psi^+$, to one of the four Bell states. After sending her entangled mode to Bob, he can recover the two bits by locally measuring the operators $\mc O_{1}$
and  $\mc O_{2}$.

Next we consider teleportation. Alice and Bob share a two mode state  $(\alpha\ket{0_A1_B}+\beta\ket{1_A0_B})$, and Alice wants to teleport her mode to Charlie.
(Fig 1.) To do that, Alice and Charlie share a resource of a fermionic e-mode
$\ket{0_A1_C}+\ket{1_A0_C}$.
We rewrite Alice's state using the Bell basis, obtaining
\begin{eqnarray}
\ket{\psi^+}_A(\alpha\ket{0_C1_B}+\beta\ket{1_C0_B}) + \\
\nonumber \ket{\psi^-}_A(\alpha\ket{0_C1_B}-\beta\ket{1_C0_B}) + \\
\nonumber \ket{\phi^+}_A(\alpha\ket{1_C1_B}+\beta\ket{0_C0_B}) + \\
\nonumber \ket{\phi^-}_A(\alpha\ket{1_C1_B}-\beta\ket{0_C0_B})
\end{eqnarray}
Now Alice measures her Bell state using the suggested Bell operators, and communicates the result to Charlie. Finally, Charlie may need to phase flip (doesn't affect the parity of the state and therefore physically achievable) or bit flip his mode (using a local ancillary mode as above).

We now turn
to illustrate  the connection  to reference frames and SiV by showing that
teleportation is possible with bosonic entanglement that is supplemented with a reference
frame that carries a Siv resource. We therefore need to consider conversion between fermionic and bosonic mode
entanglement.
\begin{theo}
A single bosonic e-mode can be converted perfectly and reversibly into a single fermionic e-mode, using a non-entangled mixed fermion two-mode state as a reference state.
\end{theo}
\begin{proof}
Let Alice and Bob share one bosonic e-mode, and one fermionic e-mode
\begin{equation}
(\ket{0_Ax_B}+\ket{1_A\underline{x}_B})
(\ket{\overline{0_A1_B}}+\ket{\overline{1_A0_B}})\ket{0_A1_B} .
\end{equation}
Here we used an upper bar sign to denote a local bosonic mode. The  index  $x$  that equals 0 or 1.
 $\underline{x}$ denotes the negate of $x$. We have assumed that Alice and Bob each have
 a local ancillary fermionic mode.

Rearranging the state to the form
\begin{eqnarray}
\ket{00\overline{0}}_A\ket{1x\overline{1}}_B+\ket{00\overline{1}}_A\ket{1x\overline{0}}_B+\\
\ket{01\overline{0}}_A\ket{1\underline{x}\overline{1}}_B+\ket{01\overline{1}}_A\ket{1\underline{x}\overline{0}}_B
\end{eqnarray}
Alice and Bob then both perform the following unitary transformation
between their two local fermionic modes and the single bosonic mode
\begin{eqnarray}
\ket{00\overline{1}}_A\Leftrightarrow\ket{11\overline{0}}_A, \tab
\ket{01\overline{1}}_A\Leftrightarrow\ket{10\overline{0}}_A \\
\ket{1x\overline{0}}_B\Leftrightarrow\ket{0\underline{x}\overline{1}}_B
\end{eqnarray}
leaving all other basis elements unchanged and resulting with
\begin{eqnarray}
\ket{00\overline{0}}_A\ket{1x\overline{1}}_B+\ket{11\overline{0}}_A\ket{0\underline{x}\overline{1}}_B+\\
\ket{01\overline{0}}_A\ket{1\underline{x}\overline{1}}_B+\ket{10\overline{0}}_A\ket{0x\overline{1}}_B
\end{eqnarray}
which is equivalent to
\begin{equation}
(\ket{0_Ax_B}+\ket{1_A\underline{x}_B})
(\ket{0_A1_B}+\ket{1_A0_B})\ket{\overline{0_A1_B}}
\end{equation}
Therefore either the state $\phi^+$ or $\psi^+$ can be used to convert bosonic entanglement to fermionic.
Since Alice and Bob need no prior knowledge of $x$ for the process to work,
they can also use a mixture of $\phi^+$ or $\psi^+$  with equal probability.
Such a mixture is not entangled as it can be expressed as a separable decomposition
of the states
\begin{equation}\label{decompose}
(\ket{0} + w\ket{1})_A(\ket{0} + w\ket{1})_B
\end{equation}
where $w$ is averaged on 1,-1 with equal probability.
\end{proof}

We comment that since the decomposition (\ref{decompose}) violates parity SSR, it cannot be physically realized in a preparation process. Therefore this state is nevertheless a non-local resource.

We also notice that the fermionic reference state remains intact
and has a role of an enabler rather then of a consumed resource. Therefore a  single reference state can be used repeatedly to convert any number of bosonic e-modes into fermionic e-modes, or vise-versa. Since bosonic mode entanglement is equivalent to ordinary non-identical e-bits,
 in the presence of a shared fermionic reference state
all known quantum information protocols can be performed on fermions.

Nevertheless, without conversion to bosonic modes, we have seen that
the basic known processes, teleportation and dense coding can be manifested with only fermionic entanglement. This holds also for distillation and dilution.
\begin{theo}
N non-perfect fermion e-modes can be converted into $N*S(\rho_A)$ fermion e-modes for large enough N.
\end{theo}
\begin{proof}
Alice begins by performing a collective measurement, in our case of the total fermion number of a  state with N copies of non-perfect fermion e-modes
\begin{equation}
(\alpha\ket{0_A1_B}+\beta\ket{1_A0_B})
\end{equation}
 The most probable outcome is $m\equiv{N}{|\beta|}^2$ and the resulting state is comprised of all the permutations of $m$ 1's in an N length vector. What is left to be done is to operate on the states with a local unitary transformation so that the final state will be a fixed fermion state times several fermion e-modes.  It can easily be shown that the number of e-modes  is $k\equiv{N}*S(\rho_A)$.

The required local unitary can be constructed as a sequence of two mode gates such as a CNOT.
However such gates violate parity SSR. Since the
local parity is not defined, adding  local ancillas
will not help since they will become entangled with the state.
The trick is then to use an ancillary e-mode. Since the  global parity of each term in the state is identical, the transformation will have the effect of locally flipping (or not) the modes of the extra e-mode. The net global effect of this on the e-mode will be identical for all terms and hence the final state of the ancillary entangled modes will remain intact.
Finally, to achieve that, Alice and Bob need to have an extra single perfect e-mode. Since an e-mode can be produced in a probabilistic process
by consuming a finite number of the pairs, the asymptotic rate remains
unchanged.
\end{proof}

A similar proof can be given for the dilution process.


After establishing that fermion entanglement can be used for ordinary quantum information tasks, we turn to show that it is basically different from boson entanglement. To be precise, we will show that a single fermionic e-mode cannot be converted to any amount of bosonic entanglement, and any amount of bosonic entanglement cannot be converted to fermion entanglement.
To begin with, we recall that since fermions states cannot contain superpositions of different \emph{global} parity states, the Hilbert space of the system must be decomposed into a direct sum of coherent parity states (following~\cite{Schuch}) $\mc H=\mc H_{-1}\bigoplus\mc H_{1}$. Moreover, every physical operator $\mc O$ must not disturb the SSR; this implies that $[\mc O,\hat P]=0$, or
\begin{equation}
\mc O = P_{1}\mc OP_{1} +  P_{-1}\mc OP_{-1}
\end{equation}
where $P_i$ are the appropriate projection operators.
We can also define  \emph{local} parity operators $\hat P_A$, $\hat P_B$ for a bipartite system. These operators obey the relation
\begin{equation}\label{eq:paritymult}
\hat P = \hat P^A \otimes \hat P^B
\end{equation}
From the above equation, it is obvious that any local operator must be of the form
\begin{equation}
\mc O_A = P^A_{1}\mc O_AP^A_{1} +  P^A_{-1}\mc O_AP^A_{-1}
\end{equation}
This equation can be extended if we allow adding ancillaries to the state.  Using Eq.(\ref{eq:paritymult}) and the fact that any ancillary system must have a definite parity, we infer that adding ancillaries can either flip or not flip the parity of the whole state. Therefore,
\begin{equation}
\mc O_A = P^A_{anc}\mc O_AP^A_{1} +  P^A_{-anc}\mc O_AP^A_{-1}\label{eq:opertor-structure}
\end{equation}
where $anc\in \{1,-1\}$. in this case we have
\begin{eqnarray}
[\mc O_A,\hat P^A]=P^A_{anc}\mc O_AP^A_{1} -  P^A_{-anc}\mc O_AP^A_{-1} - \\\nonumber anc(P^A_{anc}\mc O_AP^A_{1} -  P^A_{-anc}\mc O_AP^A_{-1}) = (1-anc)\mc O_A\hat P^A
\end{eqnarray}
deriving
\begin{equation}\label{eq:commute}
\mc O_A\hat P^A = anc\hat P^A\mc O_A
\end{equation}
meaning $\hat P^A,\mc O_A$ either commute or anti-commute.

Using the last result, we can define a monotone quantity, similar to the notion of SiV, described in~\cite{Schuch}
\begin{lem}\label{lem:monotone}
For any pure state $\phi$, the measurement of $A(\phi)\equiv{\bra{\phi}\hat P^A\ket{\phi}}^2$ can only be increased on average using LOCC, in the presence of parity SSR.
\end{lem}
\begin{proof}
The most general transformation we can perform is a local POVM. In this case, Alice brings some ancillary bits into the system, performs a unitary operation and measures the ancillary bits. i.e. the system goes from $\ket{\phi}$ to $\frac{M_i^A\ket{\phi}}{\sqrt{p_i}}$ with probability $p_i = \bra{\phi}M_i^{A\dagger}M_i^A\ket{\phi}$ and
$\sum_iM_i^{A\dagger}M_i^A=\mathbbm{1}$. Thus, after the POVM
\begin{eqnarray}
A(\phi ') = \sum_i{p_i(\frac{\bra{\phi}M_i^{A\dagger}\hat P^A M_i^A\ket{\phi}}{p_i})^2}\\ \nonumber = \sum_i{\frac{{\bra{\phi}M_i^{A\dagger} M_i^A\hat P^A\ket{\phi}}^2}{p_i}}
\end{eqnarray}
where in the last transition we used Eq. (\ref{eq:commute}) and $anc^2 = 1$. We define vectors $u_i\equiv\frac{\bra{\phi}M_i^{A\dagger} M_i^A\hat P^A\ket{\phi}}{\sqrt{pi}}$, $v_i\equiv\sqrt{p_i}$. Using Cauchy-Schwarz inequality on the last equation, we can derive that
\begin{equation}
A(\phi ') \ge {(\sum_i\bra{\phi}M_i^{A\dagger} M_i^A\hat P^A\ket{\phi}})^2 = {\bra{\phi}\hat P^A\ket{\phi}}^2 = A(\phi)
\end{equation}
\end{proof}
Using the above we can derive another important result:
\begin{lem}\label{lem:probemode}
The probability of creating a perfect fermion e-mode from an initial global state $\ket{\phi}$ using LOCC, is bounded by $1-A(\phi)$.
\end{lem}
\begin{proof}
Alice and Bob manipulate their states using a POVM and obtain a global state $\ket{\phi_i}$ with probability $p_i$. Using (\ref{lem:monotone})
\begin{equation}
A(\phi) \le \sum_{i=0}^N p_i A(\phi_i)
\end{equation}
Assuming $\ket{\phi_0}$ is the only desired state, containing a perfect fermion e-mode, so $A(\phi_0) = 0$, so the above becomes
\begin{equation}
A(\phi) \le \sum_{i=1}^N p_i A(\phi_i) \le \sum_{i=1}^N p_i = 1-p_0
\end{equation}
using $A(\phi_i)\le 1$ for any state $\ket{\phi_i}$. Taking the equalities gives the best case result
\begin{equation}
p_0 = 1 - A(\phi)
\end{equation}
\end{proof}
This proves that a system that only contains imperfect fermion e-modes cannot create a perfect fermion e-mode deterministically.
With this at hand we continue to prove the no-conversion theorem.
\begin{theo}
If Alice and Bob share only boson entanglement, it is impossible for them to create a fermion e-mode with any probability.
\end{theo}
\begin{proof}
Since the initial fermion states, in this case, are separable, Alice (and Bob) must start with a state of defined parity in order to satisfy the SSR. so we have
\begin{equation}
A(\phi)=\bra{\phi}\hat P^A\ket{\phi}^2=1
\end{equation}
Using (\ref{lem:probemode}) our best success probability is 0, so it is impossible.
\end{proof}
We now prove the other direction.
\begin{theo}
If Alice and Bob posses only a single fermion e-mode, it is impossible for them to convert it to any boson entanglement in any probability, using LOCC.
\end{theo}
\begin{proof}
Let's assume the initial fermion state is $\alpha\ket{0_A1_B}+\beta\ket{1_A0_B}$ for some $\alpha$ and $\beta$. Each local operator (including a POVM) Alice or Bob applies is limited by (\ref{eq:opertor-structure}), which means it has the following structure:
$M = \overline{\mc O_{0}}\ket{0}\bra{0}+\overline{\mc O_{1}}\ket{1}\bra{1}$
or
$M = \overline{\mc O_{0}}\ket{1}\bra{0}+\overline{\mc O_{1}}\ket{0}\bra{1}$ (considering ancillas). Applying such operators on the initial state (even if both Alice and Bob do so) cannot "mix" between the sum elements. It can only drop one of them. The situation is similar to an attempt to create an e-bit locally out of a seperated state. Of course it is not achievable.
\end{proof}

In conclusion,  we have considered some implications of  fermionic entanglement to quantum information.
While single bosonic entanglement  is well known to be equivalent to standard e-bits in all respects, parity superselection rule, restricts the set of available operation for fermionic modes. For example it forbids tests of Bell's inequalities on single copies of fermion e-modes\cite{Aharonov-Vaidman}.   Nevertheless interestingly we find that in many aspects the analogy with ordinary entanglement still holds: it is possible to teleport single fermionic modes using fermionic entanglement, and furthermore ordinary dense coding  and entanglement distillation are also deterministically possible.  Hence in a second quantized framework where "modes" replace qubits, the same set of quantum information operations are available for both fermions and bosons. We therefore conclude that from the point of view of quantum information
the mode entanglement measure suggested in \cite{Zanardi} is an operational measure which quantifies the required sources needed for deterministic teleportation or dense coding.

 Finally we find that conversion between fermion to boson entanglement is possible with the help of a non-entangled (but non-local) fermion reference frame which clearly illustrates the role of Siv resouces.  Nevertheless, within the present framework, since entangled fermion pairs are convertible to bosonic entanglement but not vise versa, fermionic entanglement seems to be a more fundamental entity.

We acknowledge the Israel science foundation grant no. 784/06, and 
German-Israeli foundation Grant no. I-857, and the European Commission under the Integrated Project QAP funded by the IST directorate as Contract Number 015848.


\end{document}